\documentclass[prl,aps,preprint]{revtex4-1}
\usepackage{graphicx}
\usepackage{epsfig}

\begin{document}

\title{Popper's Thought Experiment Reinvestigated}

\author{Chris D. Richardson}
\email{cricha5@lsu.edu}
\author{Jonathan P. Dowling}
\affiliation{Louisiana State University}

\begin{abstract}
Popper's original thought experiment probed some fundamental and subtle rules of quantum mechanics.  Two experiments have directly and indirectly tested Popper's hypothesis, but they seem to give contrasting results.  The equations governing these two experiments and Popper's thought experiment will be derived from basic quantum principles.  The experimental constants will be inputted and it will show that the two experiments agree with each other and with quantum theory.
\end{abstract}

\maketitle

\section{Introduction}
\label{sec:intro}
Karl Popper posed an interesting thought experiment in 1934 \cite{bib:popper}.  With it, he meant to question the completeness of quantum mechanics (QM).  He claimed, in the same way that Einstein, Podolsky and Rosen did \cite{bib:epr}, that the notion of quantum entanglement leads to absurd scenarios that cannot be true in real life and that an implementation of his thought experiment would not give the results that QM predicts.  Unfortunately for Popper, it has taken until recently to perform experiments that test his claims.  However, the results of the experiments do not refute QM as Popper predicted, but neither do they confirm what Popper claimed QM predicted.

In 1999 Kim and Shih implemented Popper's thought experiment in the lab \cite{bib:shih}.  The experiment, while well done with clear results, was not able to answer all questions and has instigated many interpretations \cite{bib:short}\cite{bib:sancho}\cite{bib:qureshi} of the results.  The results show some correlation between entangled photons, but not in the way that Popper thought, nor in the way a simple application of QM might predict.  A different experiment on ghost-imaging done in 1995 by Strekalov, \emph{et al.} \cite{bib:strek} sheds light on the physics behind Popper's thought experiment and the results found by Kim and Shih, but does not try to directly test Popper's thought experiment.  These two experiments are similar, but give different and unexplained results.  I will use QM to build the physics of Popper's thought experiment from the ground up and show how the results of both of these experiments agree with each other and the theory of QM.

\section{Popper's Thought Experiment}
\label{sec:popperexp}

\begin{figure*}
  \includegraphics[width=0.75\textwidth]{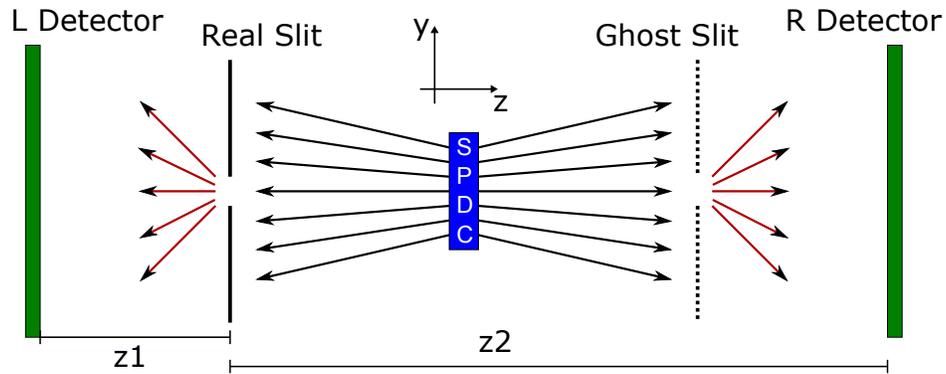}
\caption{What Popper's original thought experiment may look like using a SPDC source of photons.  The source emits a pair of momentum entangled photons in opposite directions.  The photon on the left encounters a slit that causes diffraction.  The ``ghost'' slit is the theoretical slit that Popper and others think exist due to the action of the real slit on the left.}
\label{fig:PopperGedank}
\end{figure*}

\begin{figure*}
  \includegraphics[width=0.75\textwidth]{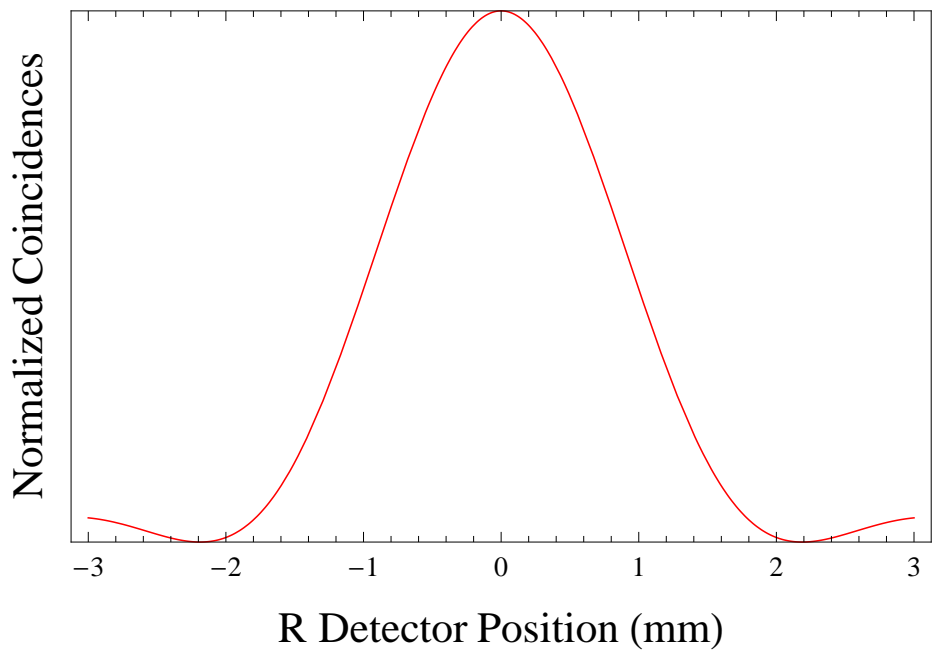}
\caption{The theoretical distribution of photons on the right that Popper thought quantum mechanics predicts.  It is a single slit interference pattern governed by the real slit width.}
\label{fig:PopperGedankData}
\end{figure*}

Popper proposed an experiment (Fig. \ref{fig:PopperGedank}) in which two photons entangled in position and momentum were sent in opposite directions \cite{bib:popper}.  The photon on the left passes through a slit.  The result of many of those photons passing through the slit would produce an interference pattern on a screen behind the slit.  This is the well-understood single slit diffraction experiment.  The action of the slit can be thought of as a measurement of the $y_L$ position of the photon.  The diffraction of the photon can be thought of as a direct consequence of Heisenberg's uncertainty principle.  Since the photon's $y_L$ position was measured, then it's momentum $p_L$ is uncertain.  Now, since we are dealing with an entangled source, Popper claimed that QM tells us that if one photon's position is measured, then the other photon's position $y_R$ is also known.  Therefore, Popper argued the momentum $p_R$ of the other photon must also be uncertain even though it did not pass through a slit.  Then, according to Popper, this photon too, when measured over many trials, should produce an identical interference pattern (Fig. \ref{fig:PopperGedankData}) compared to the photon that passes through the real slit.

Popper did not think that this would happen in an experiment.  He thought that this sort of instantaneous action at a distance was incorrect.  He argued that the diffraction of the right side photon by a ghost slit is what the theory of quantum mechanics predicts, but when the experiment was performed, no diffraction of the undisturbed photon would appear and this would therefore prove that QM is wrong or at least incomplete.  The analysis of the following two experiments below will eventually show that Popper's qualms with QM are unfounded.

\section{Kim and Shih's experiment}
\label{sec:kimshi}

\begin{figure*}
  \includegraphics[width=0.75\textwidth]{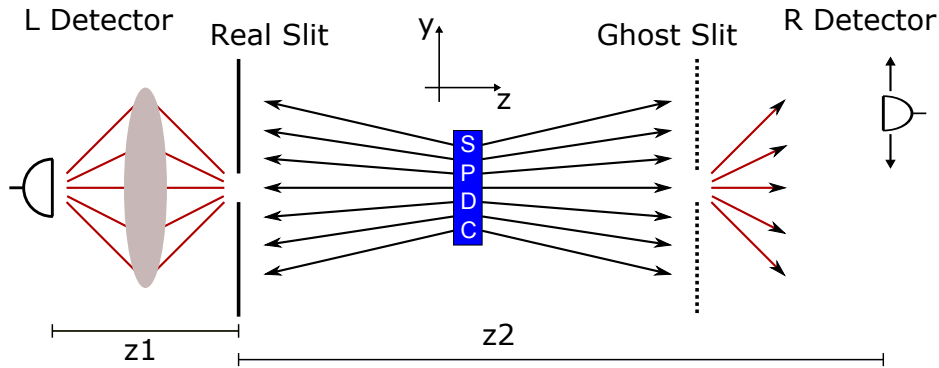}
\caption{Kim and Shih's experimental setup with a theoretical ghost slit.  The SPDC source emits a pair of momentum entangled photons in opposite directions.  The photon traveling to the left travels through a slit.  All the photons on the left that make it through the slit are collected into one fixed detector.  The detector on the right is free to scan the $y$ axis.  Only detection events in which detectors fired in coincidence are reported.}
\label{fig:Popper_Shih}
\end{figure*}

\begin{figure*}
  \includegraphics[width=0.75\textwidth]{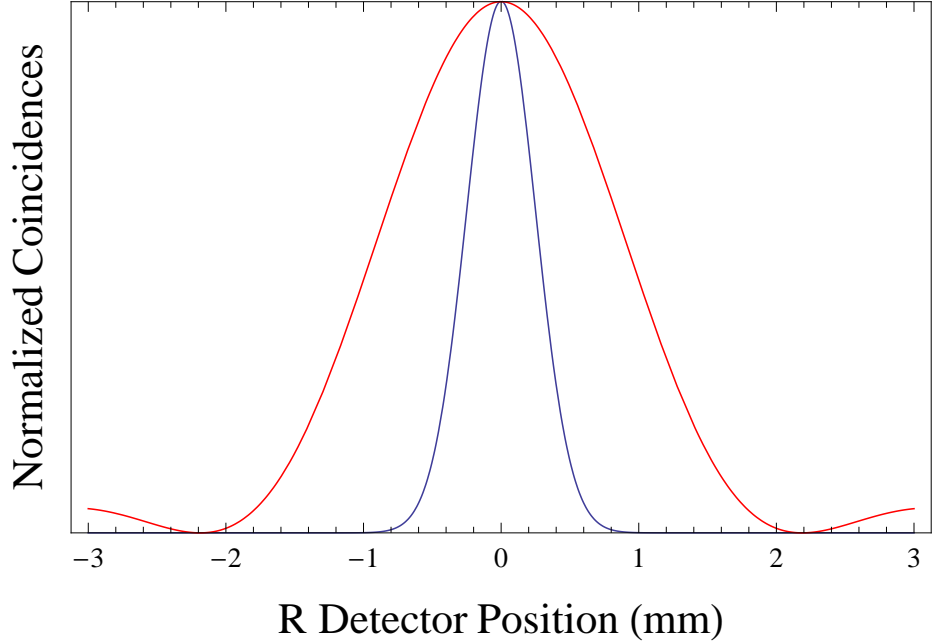}
\caption{The inside curve is a reproduction of the results of Kim and Shih's experiment.  The outside curve is the theoretical interference pattern from a single slit.  The experimental results show a momentum uncertainty less than that of single slit interference pattern.  If the photon on the right actually passes through a ghost slit of the same width of the real slit, then some violation of the uncertainty principle is suggested.}
\label{fig:Popper_Shih_Data}
\end{figure*}

Kim and Shih's experiment \cite{bib:shih} tried to directly set up and run Popper's original thought experiment.  They used a source of spontaneous parametric down conversion (SPDC) photons and sent one half through a slit and then a lens to collect all the photons that made it through the slit.  On the other side, they scanned the y-axis for coincidence counts (Fig. \ref{fig:Popper_Shih}).  The difference between this experiment and Popper's original thought experiment is that all the photons on the left side are collected through a lens and sent to one detector and the photons on the right are scanned in the $y$ axis instead of landing on some sort of screen.  The lens should have no bearing on what we see from the detector on the right, since the photon on the left still traveled through the real slit.

What Kim and Shih found is that the momentum distribution of the photon passing through the ghost slit is less than the the spread in momentum due to a real slit, and also less than the original momentum distribution from the SPDC source (Fig. \ref{fig:Popper_Shih_Data}).  This can lead to questions about the uncertainty principle.  If the photon at the ghost slit was located to a width, d = 0.16 mm, then the uncertainty principle says that it's momentum uncertainty should have increased more than the experiment shows.  The uncertainty in momentum from the experiment is $\Delta p_R \approx 3 \, \hbar \, $mm$^{-1}$, so calculating the product of the position and momentum uncertainties for the photon passing through the ghost slit gives $\Delta p_R \Delta y_R \approx \frac{\hbar}{4} < \frac{\hbar}{2}$. This seems to violate the Heisenberg Uncertainty Principle (HUP).  It is well known that while the position and uncertainty uncertainties of a single photon cannot violate HUP, the uncertainties of the addition of momentum and the difference of position commute with each other, $[\Delta(y_R-y_L),\Delta(p_R+p_L)] = 0$, and therefore need not obey HUP.  The inequality, $\Delta(y_R-y_L) \Delta(p_R+p_L) < \frac{\hbar}{2}$, can be true.  While this is a true statement, it cannot be used to get around the apparent problem brought out by the data in Kim and Shih's paper.  In the end we are measuring the momentum of each photon, not their sum.

\section{Ghost Imaging Experiment}
\label{sec:strek}

\begin{figure*}
  \includegraphics[width=0.75\textwidth]{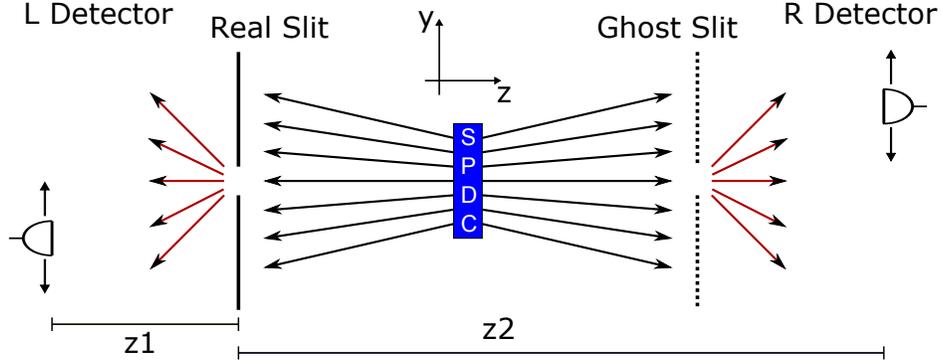}
\caption{The experimental setup of Strekalov, \emph{et al.} with a theoretical ghost slit.  The SPDC source emits a pair of momentum entangled photons in opposite directions.  The photon traveling to the left travels through a slit.  The detectors on both sides are free to scan the $y$ axis, so not all of the photons that pass through the slit on the left are collected.  Only detection events in which detectors fired in coincidence are reported.}
\label{fig:Popper_Strek}
\end{figure*}

\begin{figure*}
  \includegraphics[width=0.75\textwidth]{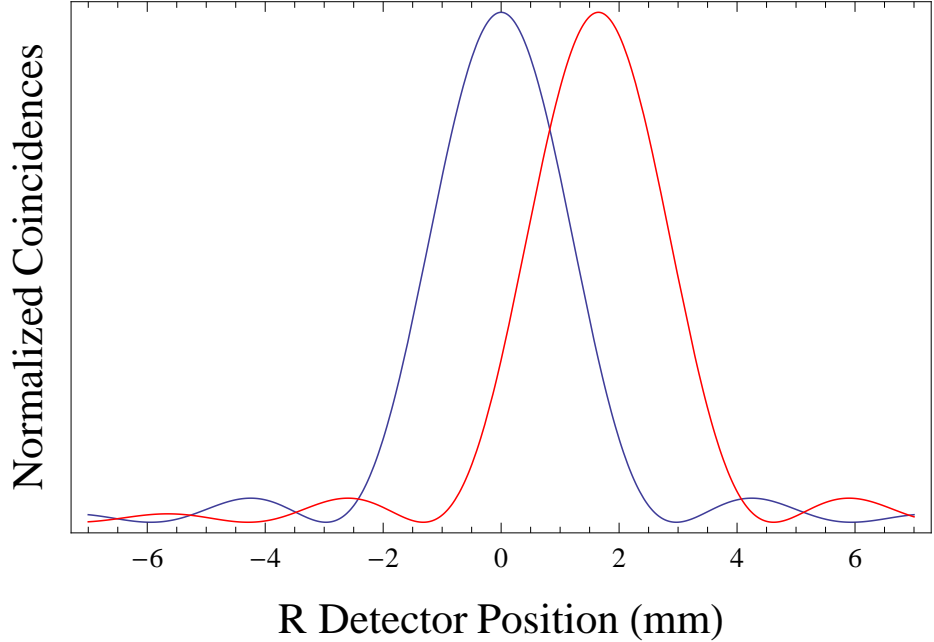}
\caption{Reproduction of experimental results of the experiment by Strekalov, \emph{et al.}  The curve centered at the origin is the scan of the $y$ axis of the detector on the right side when the detector on the left side is stationary at the origin.  The displaced curve is the scan of the $y$ axis of the detector on the right side when the detector on the left side is displaced from the origin.  Both curves are that of a single slit interference pattern, although not of the width of what would be produced by a slit of the same width as on the left side.}
\label{fig:Popper_Strek_Data}
\end{figure*}

The experiment by Strekalov, \emph{et al.} \cite{bib:strek}, was not meant to be a direct test of Popper's experiment.  It was designed to observe ghost imaging using SPDC photons.  It does, however, give a great deal of insight into Popper's experiment.

The experiment is setup in much the same way as Kim and Shih's.  A SPDC source sends entangled photons in opposite directions.  On one side there is a slit and on the other there is not.  This experiment differs from Kim and Shih's in that both sides are scanned on the y-axis.  In Kim and Shih's experiment, all the photons that passed through the slit were collected and detected, whereas in the experiment by Strekalov, \emph{et al.}, there is no lens to collect all the photons and the detector is free to scan the y-axis exactly like the detector behind the ghost slit.

The results of the Strekalov, \emph{et al.} experiment are strikingly different from Kim and Shih's.  When the detector behind the real slit is kept stationary and the detector behind the ghost slit is scanned, a single slit interference pattern was observed (Fig. \ref{fig:Popper_Strek_Data}) in coincidence counts from the detectors.  On the surface this seems to be exactly what Popper claimed QM predicts and exactly what did not happen in Kim and Shih's experiment.  Strekalov, \emph{et al.} noted two other things about the results.  When the detector behind the real slit was displaced from the center of the y-axis, the single slit interference pattern behind the ghost slit was also displaced, but in the opposite direction.  Also, the width between the fringes was dependent on the distance between the real slit and the detector behind the ghost slit, a distance they labeled as $z_2$ (Fig. \ref{fig:Popper_Strek}).

The dependence of the fringe width on the distance $z_2$ is interesting.  As Tittel, et  al. have shown \cite{bib:geneva}, entanglement is valid over extremely long distances.  If one entangled photon is measured to have a certain momentum then we know exactly the other photon's momentum, regardless of the distance between them.  It does not matter how far away two entangled particles are, their correlations are still exact.  So, the importance of this distance $z_2$ in Popper's experiment raises interesting questions that will be addressed in a later paper.  It deserves to be investigated whether this is an artifact of this particular experiment or something more fundamental.

\section{Analysis of Popper's Experiment}
\label{sec:analysis}

In the following analysis the $z$ component, the axis perpendicular to the slit, of position and momentum will be ignored since it holds no surprises and does not contribute to the analysis.  In the simplest case, plane waves with wave functions that are spread about the entire two dimensional space would be used.  It is more instructive to use a more real world starting point for the wave function from the SPDC.  We could start with a real life SPDC wave function, but the complexity either does not add to the analysis or reduces to the following wave function.  I will start with a Gaussian distribution wave function describing a momentum entangled pair of photons traveling in opposite directions.  To get the position representation of the wave function one starts by taking a Fourier transform of the momentum wave function.

\begin{eqnarray*}
	\psi(y_R,y_L) &=& \mathcal{F}(N e^{-p^2 a^2})\\
	&=& \int_{-\infty}^{\infty}{N e^{-p^2 a^2} e^{-i p (y_L-y_R)} dp}\\
	&=& N e^{\frac{-(y_L - y_R)^2}{4 a^2}}
\end{eqnarray*}

Where both photons have equal and opposite momentum $p$, $\mathcal{F}$ is a Fourier transform, $a$ is the width of the Gaussian packet, $y_L,y_R$ are the left and right components of position of the two photons and $N$ is the normalization, which will be dropped from further equations,.

The detectors are far away enough from the slit that we can use the far-field Fraunhofer approximation to find what action the slit has on the wave function after it passed through.  The slit is taken to be a box function  ($\Pi(\frac{y_L}{d}))$, or the sum of two step functions, of width $d$.  The probability amplitude in momentum space, $\Phi(p_R,p_L)$, follows.

\begin{eqnarray}
	\label{wpf}
	\phi(p_R,p_L) &=& \mathcal{F}[\psi(y_R,y_L)  \Pi(\frac{y_L}{d})] \nonumber  \\
	&=& \int_{-\frac{d}{2}}^{\frac{d}{2}}dy_L {\int_{-\infty}^{\infty}{\Psi(y_L,y_R) e^{i (y_L p_L+y_R p_R)} dy_R}} \nonumber  \\
	&=& e^{-p_R^2 a^2} \frac{\sin(\frac{d}{2}(p_L+p_R))}{(p_L+p_R)}  \\
	\Phi(p_R,p_L)&=&|\phi(p_R,p_L)|^2 \nonumber  \\
	&=& e^{-2 p_R^2 a^2} \frac{\sin(\frac{d}{2}(p_L+p_R))^2}{(p_L+p_R)^2} \label{eqn:jointp}
\end{eqnarray}

Where $p_L$ and $p_R$ are the inferred momentum and are functions of the slit width, wavelength, distance from the slit and position of the detector.  Note that the presence of a Gaussian in equation (\ref{eqn:jointp}), with the original SPDC source beam's width, shows that the spread in momentum due to a ghost slit can never be greater than the original momentum spread of the SPDC source.  If this were not so, then one could signal faster than light (FTL) just by changing the real slit width and watching for a $y$ momentum spread change in the right side detector.  It should also be noted that only when an observer has the \emph{coincidence} information from both sides of the experiment can any correlation can be seen.  This can be shown by finding the marginal probability distributions.  These show the probability densities for an observer on one side when they have no information about the other side.

\begin{eqnarray}
	\Phi(p_L) &\approx & \int_{-\infty}^{\infty}\delta(p_R) \frac{\sin(\frac{d}{2}(p_L+p_R))^2}{(p_L+p_R)^2} dp_R \nonumber  \\
	&=& \frac{\sin(\frac{d}{2}p_L)^2}{p_L^2} \label{eqn:margpL}\\
	\Phi(p_R) & = & \int_{-\infty}^{\infty}{\Phi(p_R,p_L) dp_L} = e^{-2 p_R^2 a^2} \label{eqn:margpR} 
\end{eqnarray}

The marginal distribution on the left, [eqn. \ref{eqn:margpL}], can be found by assuming the beam width, $a$, is wide compared to the slit, $a > d$.  In this case the Gaussian in the joint distribution [eqn. \ref{eqn:jointp}] can be approximated as a Dirac delta function.  After integrating the joint distribution [eqn. \ref{eqn:jointp}] over $p_R$, the marginal distribution [eqn. \ref{eqn:margpL}] gives us the single slit interference pattern that we expect from a photon going through a single slit.  So with no information from the right side detector we get exactly what we thought we should get, single slit diffraction.  To get the marginal distribution on the right side  [eqn. \ref{eqn:margpR}], we integrate the joint distribution [eqn. \ref{eqn:jointp}] over $p_L$.  Only a Gaussian with the original width remains.  So, without any information from the left side, we are left with what was originally emitted from the SPDC source.  Only in coincidence counts will any interference pattern be seen, and since the coincidence counts are transmitted over a classical channel, we are again safe from any FTL signaling.

These equations have the correct form for the results of the experiment by Strekalov, \emph{et al.}, but there is one problem.  Strekalov, \emph{et al.} noticed that the probability distribution is dependent on the aforementioned distance $z_2$ from the real slit to the detector behind the ghost slit.  But if we stay with the notion of a ghost slit, then one would assume the probability distribution behind the ghost slit would only depend on the distance from the ghost slit to the detector and on the width of the ghost slit.  This is not the case.

\section{Resolution of the Two Experiments}
\label{sec:resolution}

\subsection{Bi-Photon}
\label{sec:biphoton}

The problem comes from having assumed that the ghost slit is placed at a certain position; that some measurement of the $z$ position has been made.  If this were so, then we could justify the presence of a ghost slit at the position opposite that of the real slit.  For, as we know, if the position of one photon is measured, then the position of the other is known.  But, the $z$ axis components of the wave function has in no way, weak or otherwise, been measured or restricted.  No ghost slit can been made by only measuring the y-component of the wave function at the real slit.  So, we should not use the location of a ghost slit as a reference point. This leaves the only point of reference for the partial collapse of the initial wave function to be the position of the real slit.  So all momentum changes must be referenced from that point.  This is where the notion of two entangled photons traveling in opposite directions breaks down and is what has lead Shih and others to refer to this entangled state as a ``bi-photon''.  The measurement due to the slit on the left side can not be thought to measure only one photon and then through entanglement have an effect on the other photon.  The action of the real slit must be understood to affect the bi-photon wave function.

\begin{figure*}
  \includegraphics[width=0.75\textwidth]{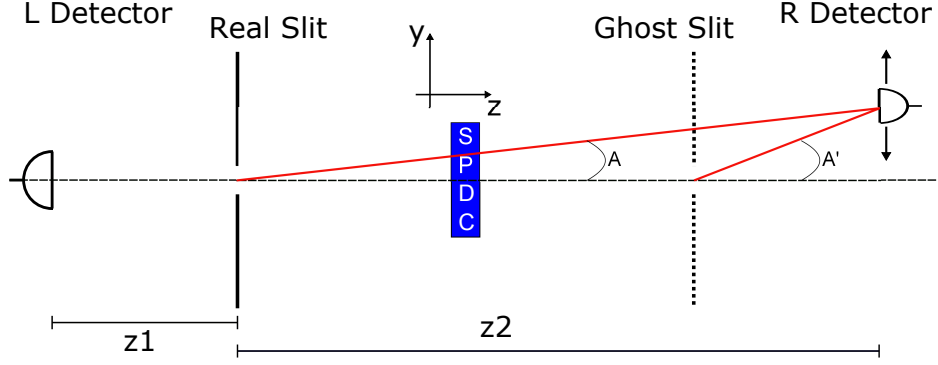}
\caption{The angle of diffraction changes depending on which reference frame you choose.  In this case, the angle from the reference frame of the real slit, $A$, is smaller than the angle, $A'$, taken from the reference frame of the ghost slit.}
\label{fig:Popper_Angles}
\end{figure*}

Since we are abandoning the notion of a ghost slit, the distance between the ghost slit and the detector on the right is not a number that can be used.  The momentum vectors of both sides need to have their origin at the real slit.  Figure \ref{fig:Popper_Angles} shows that the angle, $A'$, referenced from the ghost slit will be larger than the angle, $A$, referenced from the real slit.  The translation has the effect on the right side of extending the distance $z_1$ that is normally seen in the distributions of single slit interference patterns to the much larger distance $z_2$ that runs from the plane of the real slit to the plane of the detector behind the ghost slit.  Then the probability amplitude at the detector on the right is:

\begin{eqnarray}
	\Phi(y_R,y_L) & = & e^{-\frac{8 \pi^2 y_R^2 a^2}{\lambda^2 z_2^2}} \frac{\sin(\frac{\pi d}{\lambda}(\frac{y_L}{z_1}+\frac{y_R}{z_2}))^2}{(\frac{y_L}{z_1}+\frac{y_R}{z_2})^2} \nonumber  \\
	& \approx & \frac{\sin(\frac{\pi d}{\lambda}(\frac{y_L}{z_1}+\frac{y_R}{z_2}))^2}{(\frac{y_L}{z_1}+\frac{y_R}{z_2})^2} \label{eqn:addzcond}
\end{eqnarray}

\subsection{The two Experiments Resolved}
\label{sec:resolved}

\begin{figure*}
  \includegraphics[width=0.75\textwidth]{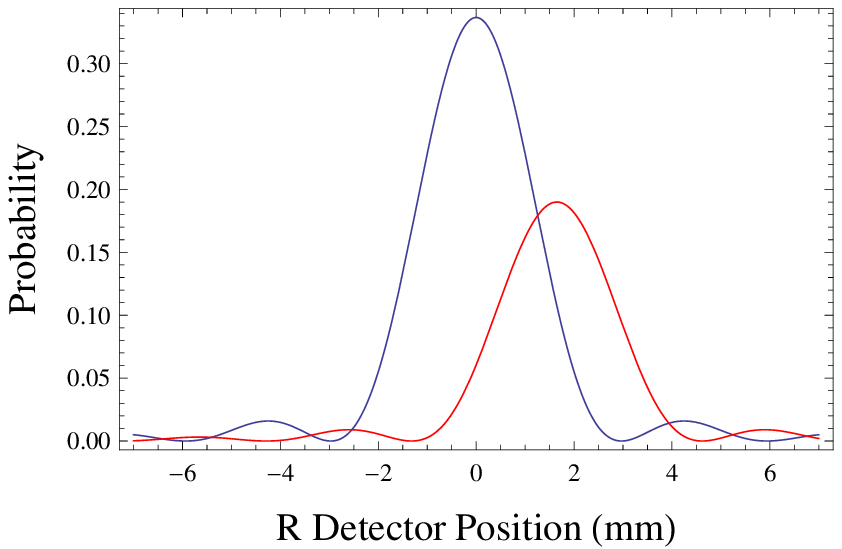}
\caption{Theoretical probabilities for detecting photons at the detector on the right for the experiment by Strekalov, \emph{et al.}  The curve centered at the origin is the theoretical probability distribution of intensity on the $y$ axis on the right side when the detector on the left side is stationary at the origin.  The displaced curve is the theoretical probability distribution of intensity on the $y$ axis of the detector on the right side when the detector on the left side is displaced from the origin.  Both curves are that of a single slit interference pattern.  The width of the pattern is governed by the distance between the real slit on the left to the detector on the right.  These theoretical results match those given in experiment.}
\label{fig:Popper_ForReals_Strek_Data}
\end{figure*}

Equation \ref{eqn:addzcond} gives us all the results that the experiment by Strekalov, \emph{et al.} measured (Fig. \ref{fig:Popper_ForReals_Strek_Data}).  Note that this theoretical figure is normalized by probability rather than coincidence counts as is the case for the figure of experimental results (Fig. \ref{fig:Popper_Strek_Data}).  One can see that the probability for detecting photons when the left detector is displaced from the large central peak is much less.

\begin{figure*}
  \includegraphics[width=0.75\textwidth]{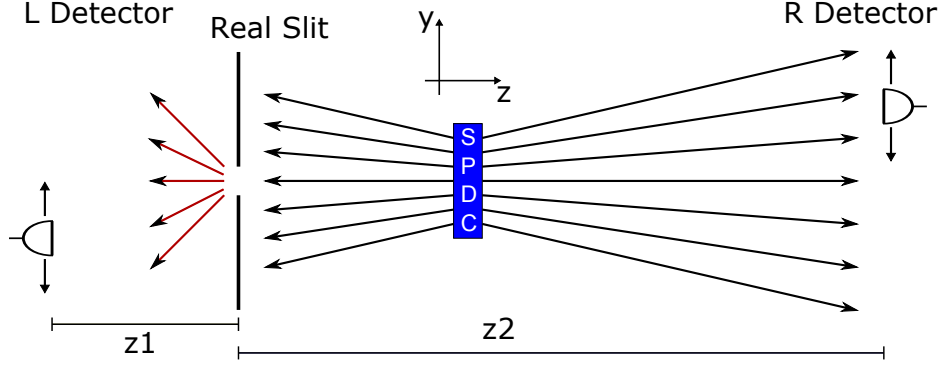}
\caption{The experimental setup of Strekalov, \emph{et al.} without the notion of a ghost slit.  The SPDC source emits a pair of momentum entangled photons in opposite directions.  The left part of the bi-photon passes through a slit.  This has a measurable effect on the right side of the bi-photon, but the notion of a ghost slit on the right side is discarded.  The detectors on both sides are free to scan the $y$ axis, so only some of the photons that pass through the slit on the left are collected.  Only detection events in which detectors fired in coincidence are reported.}
\label{fig:Popper_ForReals_Strek}
\end{figure*}

Kim and Shih's results are a little more mysterious.  Kim and Shih found that the width of the distribution after the ghost slit was less then the original beam width and less than the width of the single slit interference pattern of a real slit at the ghost slit's position.  This does not agree with the assumption that the ghost slit will cause an increase in the momentum spread.  It also does not seem to agree with Strekelov's result that indicates the width will be that of single slit interference pattern, augmented by the distance $z_2$ between the detector behind the ghost slit and the real slit.  The key to understanding Kim and Shih's results is to realize that the only substantial difference between their experiment and Strekelov's experiment is that all of the photons were collected after the real slit.  They, in effect, took a weighted average of all the photons that passed through the slit.  With the weighting function, $W(y_R,y_L)$, being that of a single slit interference pattern.  So if we multiply the probability amplitude of joint detection by the weighting function and then integrate over all the detected photons that came through the left slit, we get the counting rate, $R(y_R)$ that Kim and Shih measured in their experiment:

\begin{eqnarray*}
R(y_R) & = & \int_{-\infty}^{\infty} W(y_R,y_L) \Phi(y_R,y_L) dy_L \nonumber  \\
	& = & \int_{-\infty}^{\infty} e^{-\frac{8 \pi^2 y_R^2 a^2}{\lambda^2 z_2^2}} \frac{\sin(\frac{\pi d}{\lambda}(\frac{y_L}{z_1}))^2}{(\frac{y_L}{z_1})^2} \frac{\sin(\frac{\pi d}{\lambda}(\frac{y_L}{z_1}+\frac{y_R}{z_2}))^2}{(\frac{y_L}{z_1}+\frac{y_R}{z_2})^2} dy_L \nonumber  \\
	&=& \frac{e^{\frac{8 \pi^2 y_R^2 a^2}{\lambda^2 z_2^2}} (\frac{\pi y_R d}{z_2 \lambda}-\sin[\frac{\pi y_R d}{z_2 \lambda}])}{y_R^3} \label{eqn:ShihProb}
\end{eqnarray*}

\begin{figure*}
  \includegraphics[width=0.75\textwidth]{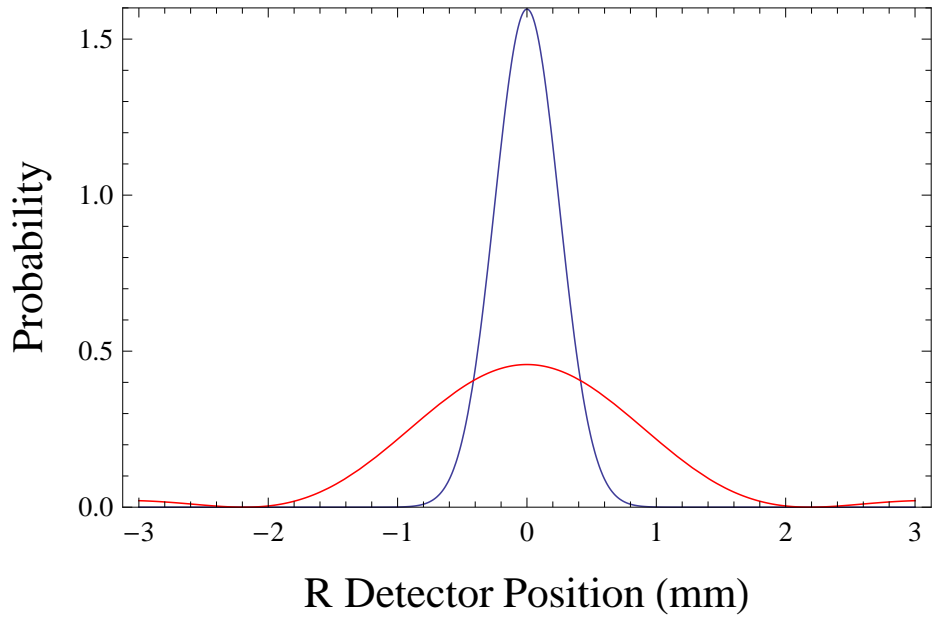}
\caption{The inside curve is the theoretical probability distribution of intensity on the $y$ axis on the right side for the experiment by Kim and Shih.  The outside curve is the probability distribution of intensity for single slit diffraction.  These theoretical results match those given in experiment.}
\label{fig:Popper_ForReals_Shih_Data}
\end{figure*}

And if we insert the values that Kim and Shih used in their experiment we get figure \ref{fig:Popper_ForReals_Shih_Data};  where the red line on the outside is what one would get for single slit diffraction at a real slit and the blue line on the inside is what Kim and Shih got on the detector on the right, behind the ghost slit.

\begin{figure*}
  \includegraphics[width=0.75\textwidth]{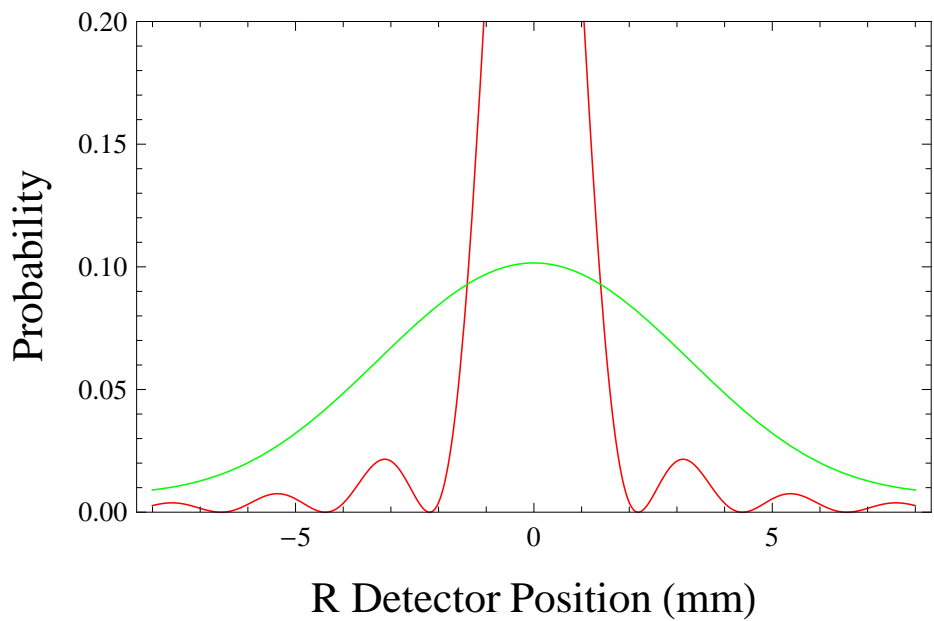}
\caption{The outside curve is the theoretical probability distribution of intensity on the $y$ axis on the right side for the experiment by Kim and Shih if the beam width is taken to go to infinity.  The inside curve is the probability distribution of intensity for single slit diffraction.}
\label{fig:Popper_ForReals_Shih_Data_Limit}
\end{figure*}

If one were to assume a ghost slit were present, that would mean that the position of the photon on the right was known to be in the region of the slit width, but it's momentum uncertainty is less than what would be allowed by Heisenberg's uncertainty principle.  But as I have said, no measurement of the $z$ position was taken and there is no ghost slit, therefore no violation of the uncertainty principle.  What we end up with in Kim and Shih's experiment is the sum of all those displaced single slit interference patterns noticed in Strekelov's experiment weighted by the probability of the interference pattern of the real single slit.  The intensity distribution is something that looks like a Gaussian with a width dependent on the distance between the real slit and the detector behind the ghost slit, the beam width, the wavelength and the slit width.  The edges of this curve are greatly attenuated by the width of the beam.  If one were to take the limit of $R(y_R)$ as the beam width gets large, $a \rightarrow 0$, one could see from the rescaled figure \ref{fig:Popper_ForReals_Shih_Data_Limit} that the curve becomes much broader.

\begin{figure*}
  \includegraphics[width=0.75\textwidth]{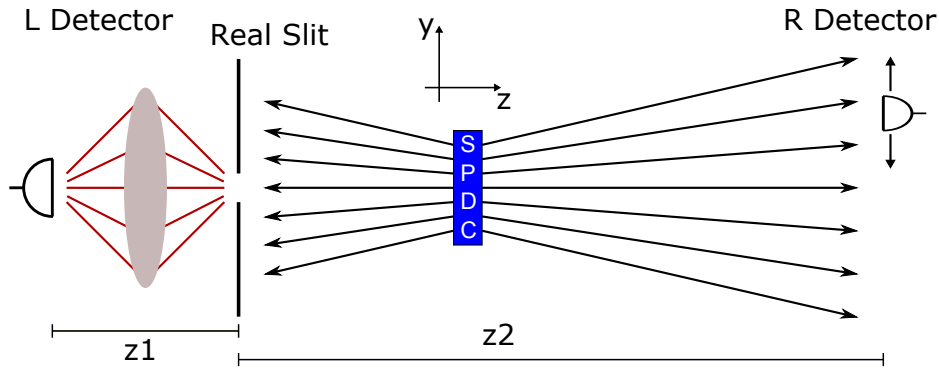}
\caption{Kim and Shih's experimental setup without the notion of a ghost slit.  The SPDC source emits a pair of momentum entangled photons in opposite directions.  The left part of the bi-photon passes through a slit.  This has a measurable effect on the right side of the bi-photon, but the notion of a ghost slit on the right side is discarded.  All the photons on the left that make it through the slit are collected into one fixed detector.  The detector on the right is free to scan the $y$ axis.  Only detection events in which detectors fired in coincidence are reported.}
\label{fig:Popper_ForReals_Shih}
\end{figure*}

\section{Conclusion}
\label{sec:conclusion}

The results from both experiments flow from the same fundamentals and do not commit any egregious crimes against QM such as FTL signaling and HUP violation.  The notion of a ghost slit never comes out of the analysis, it is an artificial assumption forced onto the experiment.  I agree that the ghost slit sounds right and I imposed it when first looking into Popper's experiment, but the data from two well run experiments and the preceding analysis forces it to be abandoned. The fact that the results of Popper's thought experiment were never correctly calculated in the seventy plus years since it was proposed illustrates just how unintuitive and strange QM can be.  Unlike Popper, we are fortunate to have decades of research showing just how correct QM is, so instead of using this experiment as a test of QM it should be viewed as an interesting and subtle application of quantum mechanical fundamentals that gives up a deeper understanding of QM.  Popper argued that QM was incomplete and that an experiment would not show spooky action at a distance at a ghost slit.  He was right about the ghost slit, but wrong about QM.  Assuming that the photon on the right that does not pass through the slit is still measured at a ghost slit is incorrect.  No measurement of the $z$ component of the wave function was made, so placing a ghost slit at a specific place on the $z$ axis should not be done.  Being able to show that the two different experimental results come from the same QM principles gives credence to this conclusion.  

\begin{table}
\caption{Table of experimental constants used by Strekalov, \emph{et al.}, and Kim and Shih.}
\label{tbl:constants}
\begin{tabular}{|l|l|l|} 
\hline 
\multicolumn{3}{|c|}{Experimental Constants} \\ 
\hline 
& Kim and Shih & Strekalov, \emph{et al.} \\
\hline
Wavelength ($\lambda$)& 0.0007 mm & 0.000702 mm \\
\hline
Slit Width ($d$) & 0.16 mm & 0.4 mm \\ 
\hline 
$z_1$ & 500 mm & 1000 mm \\ 
\hline 
$z_2$ & 1500 mm & 1650 mm \\ 
\hline 
Beam Width ($\frac{1}{a}$) & 3 mm & not given \\ 
\hline
\end{tabular}
\end{table}

\begin{acknowledgments}
I would like to thank the Foundational Questions Institute for their generous support.
\end{acknowledgments}

\end{document}